\documentclass[
10pt, 
letter, 
oneside, 
headinclude,footinclude, 
]{scrartcl}

\usepackage{amsmath,amssymb,amsthm} 
\usepackage{lyquor} 

\usepackage[T1]{fontenc} 
\usepackage[utf8]{inputenc} 
\usepackage{minted}
\usepackage[margin=52pt,columnsep=0.33in, head=1in]{geometry}
\usepackage[table]{xcolor} 
\usepackage{pifont}       
\usepackage{array}        

\graphicspath{{figures/}} 

\definecolor{checkgreen}{HTML}{4CAF50}  
\definecolor{crossred}{HTML}{E53E3E}    
\definecolor{cautionorange}{HTML}{FF8C00}  
\definecolor{titlecolor}{HTML}{3a12ff}  
\newcommand{\cmark}{\textcolor{checkgreen}{\ding{51}}}%
\newcommand{\xmark}{\textcolor{crossred}{\ding{55}}}%
\newcommand{\ymark}{\textcolor{cautionorange}{\ding{108}}}%

\usepackage{enumitem} 

\usepackage{lipsum} 

\usepackage{subfig} 

\usepackage{varioref} 

\theoremstyle{definition} 

\theoremstyle{plain} 

\theoremstyle{remark} 

\hypersetup{
pdftitle={LYte-QUORum: Off-Chain Ready Smart Contract Hosted with Choice}, 
pdfauthor={Lyquor Labs}, 
pdfsubject={}, 
pdfkeywords={}, 
pdfcreator={pdfLaTeX}, 
pdfproducer={LaTeX} 
}

\title{\protect\raisebox{-0.3ex}{\includegraphics[height=0.3in]{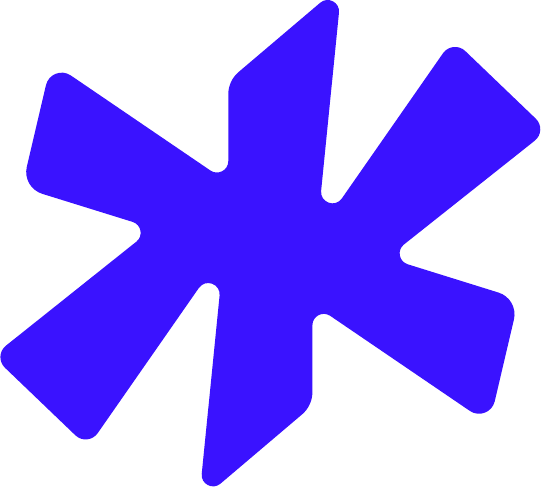}}\hspace{0.5em}{\color{titlecolor}\textsc{LY}}\textsc{te} {\color{titlecolor}\textsc{QUOR}}\textsc{um}} 
\subtitle{\subtitletheme{Off-Chain Ready Smart Contract Hosted with Choice}} 
\author{\\[0.05in]\large\sffamily Hao Hao, Dahlia Malkhi, Maofan Yin, Lizan Zhou\\\large\subtitletheme{Lyquor Labs, Inc.}} 
\date{} 

\begin{document}

\pagestyle{scrheadings} 

\newcommand{\TODO}{\textcolor[rgb]{1,0,0}{(TODO) }}
\newcommand{\ted}[1]{{\color{blue}Ted: #1}}
\newcommand{\comment}[1]{}

\maketitle 

\setcounter{tocdepth}{2} 











\section{Rethinking Blockchain Infrastructure}
Lyquor is a decentralized platform for \emph{off-chain ready} smart contracts, hosted with choice. It unifies on-chain and off-chain logic through a single programming model, powered by:

\begin{itemize}
    \item Smart coordination between nodes for selective hosting,
    \item Seamless state preservation across executions without overhead, and
    \item Aggregation-style composability of off-chain computation.
\end{itemize}

DeFi reveals the architectural limits of today's blockchains. To improve pricing and performance, and to reduce MEV, protocols like UniswapX~\cite{uniswapx} and CoWSwap~\cite{cowswap} \emph{increasingly rely on} off-chain solvers and wallet-side routing to match trades based on user intents, moving execution logic outside the contract. Others, like Hyperliquid~\cite{hyperliquid}, deploy custom chains to gain speed and predictability, at the cost of fragmentation. These trends highlight a growing need for \emph{dedicated execution environments} tailored to each service, yet without giving up composability or scalability. Lyquor addresses this need by enabling Selective Hosting (Section~\ref{sec:selective-hosting}) and off-chain capable contracts (Section~\ref{sec:offchain-computation}), allowing services to scale independently while staying integrated within a unified system for liquidity and coordination.

The same limitations appear in emerging sectors outside finance. Recent experiments in "AI+Crypto", SocialFi, GameFi categories aim to combine decentralized infrastructure with application logic. While directionally promising, their approach suffers from a critical flaw. Only the underlying \emph{data} is verified on chain, while the \emph{interpretation} of that data, the logic that powers services and drives key decisions, is carried out off-chain in a centralized and opaque way.

This is not about moving all computation back on chain. In most existing designs, the parts of the system that actually determine user outcomes, such as rankings, payouts, GPU orchestration or game cluster management, could be made directly driven by on-chain contracts. Other application-specific components could have been designed more deliberately to make off-chain computation verifiable. But they are often not engineered that way, due to the lack of a unified and flexible framework. The result is an illusion of decentralization, where tokens are issued with value not directly tied to the approach.

What is needed instead is a \emph{rethinking} of blockchain architecture itself:

\begin{itemize}
    \item Off-chain behavior should be seamlessly driven by on-chain state.
    \item Off-chain execution should be secure and resilient, even if it does not run on chain.
    \item We should eliminate the gap between on-chain and off-chain, both in performance and in usability.
\end{itemize}

While powerful for verifying a fixed sequence of operations, Zero Knowledge proofs are fundamentally designed to optimize \emph{replication}, not \emph{collaboration}. They allow one party to perform a computation and others to verify it without repeating the work. But they cannot address the broader challenge of distributed execution, where specialized nodes contribute different parts of a computation over a final result. A complete system should enable both replication and collaboration.

Therefore, Lyquor is designed from the ground up to meet these needs.

\paragraph{Fate-Constrained Ordering (FCO)} allows nodes to host contracts selectively without losing Layer-1 grade composability, enabling flexible off-chain use cases. (Section~\ref{sec:fco})
\paragraph{Direct Memory Architecture (DMA)} eliminates the gap between on-chain and off-chain execution, enabling fast, seamless computation across both domains. (Section~\ref{sec:dma})
\paragraph{Universal Procedure Call (UPC)} extends composability beyond replication (on-chain), enabling collaboration across independently hosted services (off-chain). (Section~\ref{sec:upc})

Together, these innovations form the foundation for a service-native decentralized infrastructure, built to move beyond the limits of traditional blockchains.

While the idea of decentralization may not matter to every end user, \emph{the outcomes it enables do.} Open systems offer better pricing, broader choice, and more self-service experiences. The same applies to cloud services: a new generation of automated, self-governed, easily deployable infrastructure is needed, one that reduces friction without reintroducing central points of control.

\subsection{From Chains to Services}
Traditional blockchains, whether Layer-1~\cite{layer1}, Layer-2~\cite{layer2}, or Data Availability~\cite{dataavailability} Layer, are built around the \emph{same} core assumption: nodes run chains, and \emph{every} node is expected to execute \emph{every} transaction for \emph{every} contract.

\begin{figure}[h]
\centering
    \includegraphics[width=0.6\textwidth]{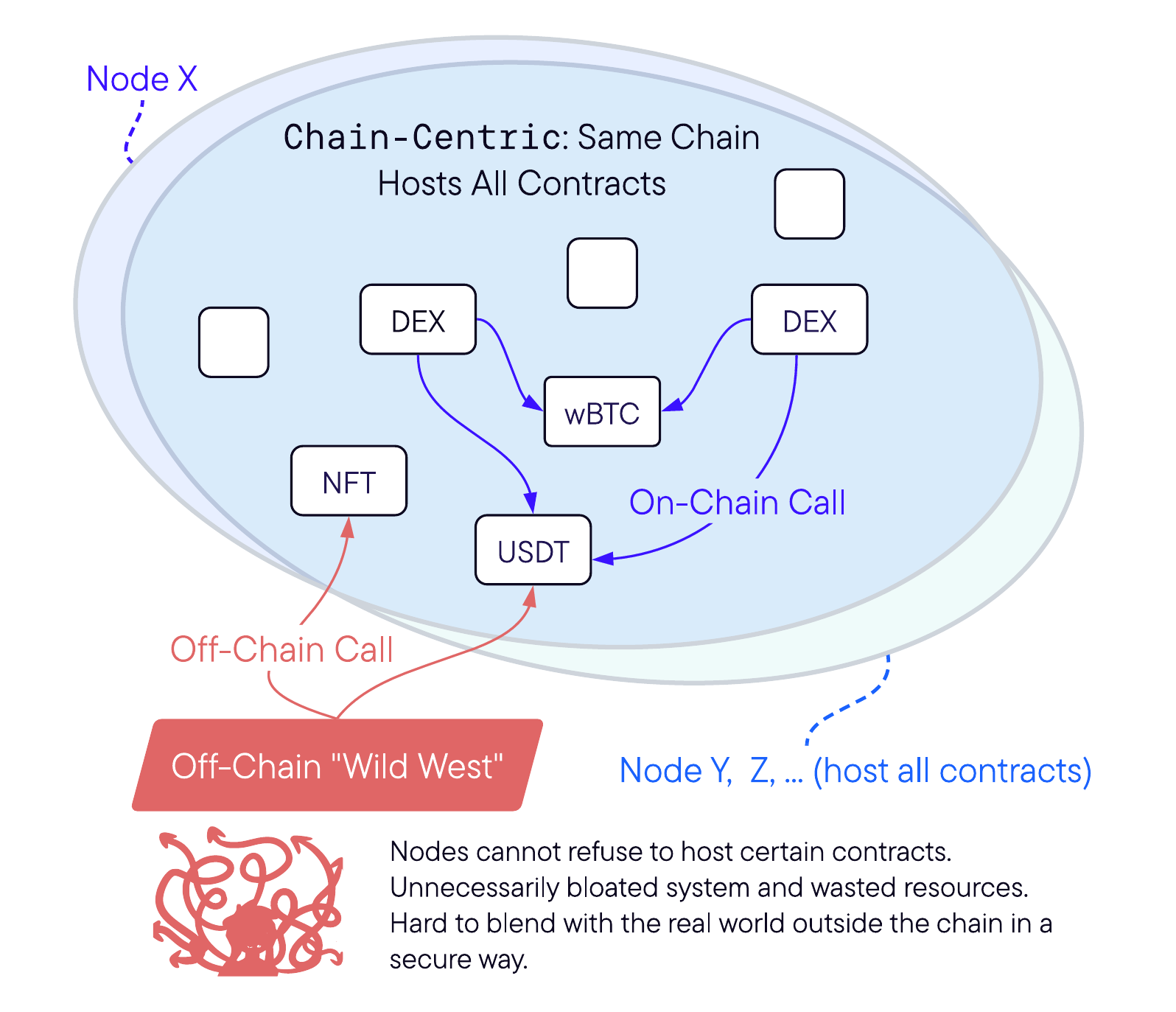}
    \caption{Traditional chain-centric architecture has some long-standing, more fundamental issues.}
    \label{fig:chain-centric}
\end{figure}

This model is increasingly unsustainable because:

\begin{itemize}
    \item Nodes are forced to carry unrelated workloads.
    \item Gas pricing becomes over-conservative, driven by worst-case assumptions, since every node must execute every transaction for every contract.
    \item Developers and node operators remain decoupled, with no direct incentive loop between service quality and infrastructure support.
    \item Attempts at scaling through subnets, parachains, or chain sharding often introduce \emph{fragmentation}, sacrificing \emph{global composability} for throughput.
\end{itemize}

Lyquor introduces a new foundation: the \emph{Service-Centric} model.

In this model, nodes no longer run chains, they run \emph{services}. Execution happens per service, not per chain. Each service maintains its own structured, persistent state and can incorporate off-chain logic natively.

We call these services Lyquids (Section~\ref{sec:lyquid}): \emph{unified, off-chain capable} smart contracts written using a lightweight Rust-based programming model. Developers write logic in idiomatic Rust, enriched with simple macros for declaring state and defining callable functions. There is no new language to learn, just durable, composable services built for decentralized hosting.

Unlike other Rust-based (WASM) smart contract frameworks, Lyquid is not more complex than Solidity~\cite{solidity}. It's often simpler. The abstractions closely mirror what developers are already used to, but without the boilerplate or manual storage management. Common patterns like value transfer and information mapping are just as concise (Section~\ref{sec:eth-friendly}) in Lyquid, while enabling zero-overhead state access and powerful off-chain logic. Lyquids also integrate \emph{seamlessly with the Ethereum API}, allowing users to interact with them using existing Ethereum tooling for \emph{both} on-chain and off-chain computation.

This service-native model enables a more efficient, modular, and incentive-aligned infrastructure, where nodes and services can form a healthy computation market without sacrificing composability or scalability.

\begin{figure}[h]
\centering
    \includegraphics[width=0.6\textwidth]{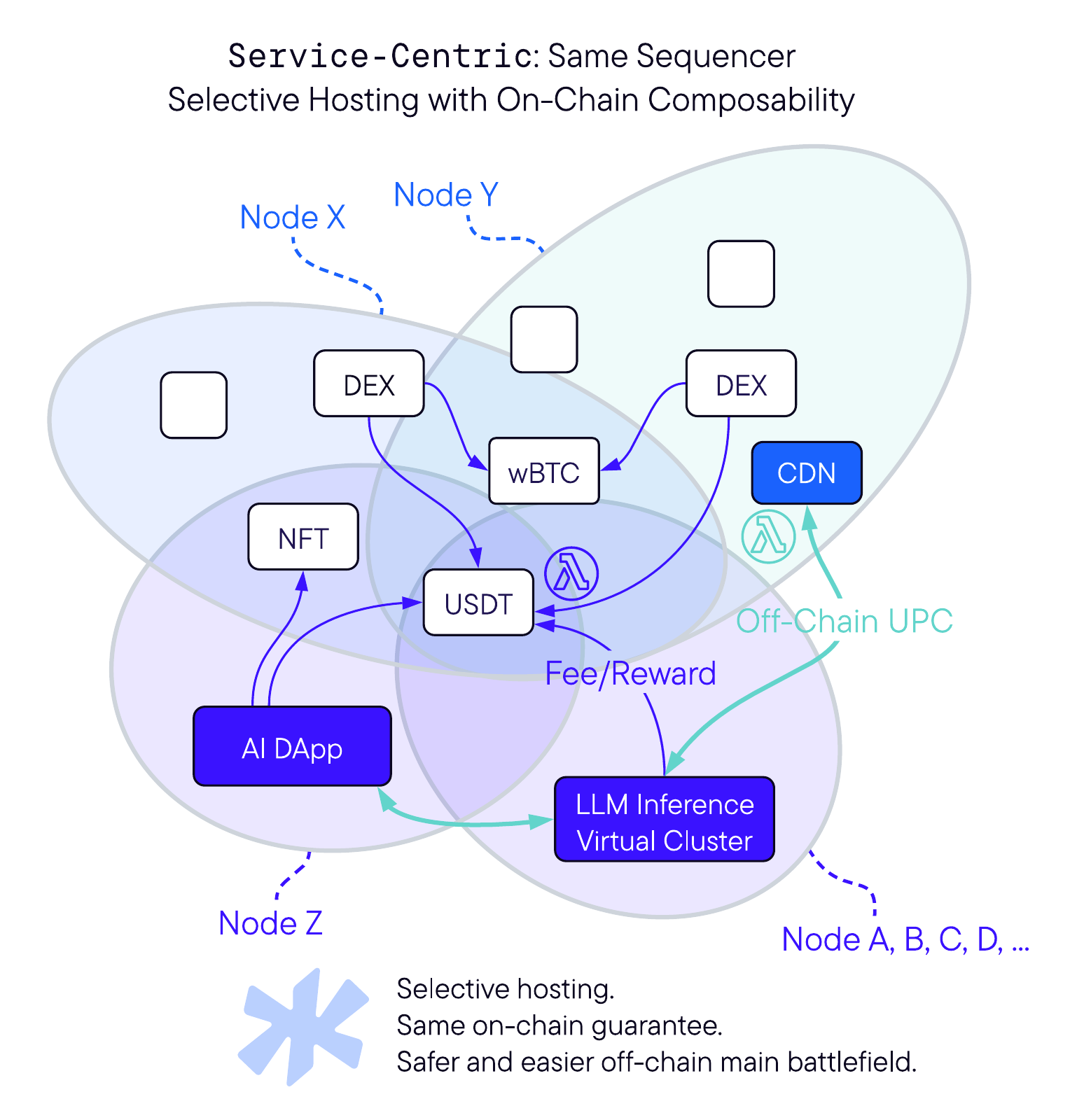}
    \caption{Lyquor's service-centric architecture where nodes run services (Lyquids) rather than chains, enabling selective hosting and specialized execution.}
    \label{fig:service-centric}
\end{figure}

\subsection{Selective Hosting with Composability}
\label{sec:selective-hosting}
In Lyquor, nodes can \emph{choose} which Lyquids to host. This alone unlocks a new execution economy:
\begin{itemize}
    \item Nodes run only the workloads that matter to them.
    \item Lyquids can offer \emph{custom incentives} to attract nodes with specialized infrastructure (e.g., GPU, storage, low-latency compute).
    \item Developers are no longer confined to one-size-fits-all fee models. They can build custom billing or credit logic (Section~\ref{sec:cloud-fee-model}).
\end{itemize}

This creates an organic, \emph{market-driven} ecosystem:
\begin{itemize}
    \item High-quality services grow stronger, attracting more nodes.
    \item Low-quality or toy Lyquids naturally fade. No protocol-level curation needed.
\end{itemize}

For example, a Lyquid offering AI inference might only be viable on GPU-enabled nodes, and another Lyquid may only be hosted by a group of nodes capable of ZK proof generation. With Lyquor, this works natively. Those nodes can join the "virtual cluster" to host that Lyquid without burdening others, and the Lyquid can reward them accordingly.

But this flexibility raises a fundamental question: how can composability be preserved without requiring every node to replicate every service's state, as in a monolithic Layer-1?

Most scalability solutions, such as subnets and sharding, solve this by partitioning the system. In doing so, they compromise the ability to coordinate atomically across services, and fragment the unified execution environment that platforms like Ethereum provide.

Lyquor avoids this tradeoff through Fate-Constrained Ordering (FCO) (Section~\ref{sec:fco}). FCO decouples execution from ordering by establishing a global causal order of contract calls before any execution takes place. The ordering described here applies to the globally consistent portion of execution. The other form, off-chain logic, is handled differently and discussed in a later section (Section~\ref{sec:offchain-computation}). This sealed timeline is produced efficiently through a lightweight sequencing layer, and every node, even if executing only a partial view, must respect the same global fate. In effect, the system's fate is already written, like a play script. Each node is an actor, not improvising. Instead, it unravels its part of the story as written. Its role is progressively discovered from the globally sequenced timeline.

The result is a platform that retains the determinism and global integrity of a monolithic Layer-1, while enabling modular, service-centric execution. With FCO, composability scales without requiring full replication.

This stands in contrast to how sequencing works in typical Layer-2 systems:

\begin{itemize}
    \item In a Layer-2, the sequencer batches transactions off-chain, executes them locally, and then eventually \emph{pushes the results back to the Layer-1 chain} for settlement.
    \item A Layer-2 is still fundamentally chain-centric. The execution is a temporary staging phase, but \emph{the main battlefield remains the Layer-1 chain} where liquidity lives, and where finality must be posted.
    \item Unlike a Layer-2, where global ordering is composed at settlement, Lyquor determines the causal order up front. Nodes simply follow the parts relevant to them rather than inventing the story as they go.
    \item Lyquor's sequencing is akin to shared rollup sequencing, in that it establishes a global ordering across autonomous chains, though in Lyquor, this applies to Lyquids rather than chains. Lyquids are not Layer-2 chains. They do not settle on a Layer-1. They settle independently.
\end{itemize}

Lyquor's sequencer is \emph{not a staging layer for another chain}. It is a coordination fabric for service-centric execution. The sequencer does not execute transactions; it only enforces global order across Lyquid calls, ensuring that composability is preserved even when execution is modular and decentralized.

While Lyquor \emph{does have a chain}, it is not responsible for performing the computation. It writes the script, but does not act it out. This change in architecture does not sacrifice the guarantees of a Layer-1. With FCO, Lyquor uses an existing consensus protocol to provide the global consistency and neutrality expected from a Layer-1, without inheriting its architectural constraints.

This enables seamless coordination across services without bridges or loss of atomicity, and allows parallel execution when Lyquids operate independently. As a result, composability scales without the burden of full replication.

\subsection{Organic Growth from Base Availability}
\label{sec:base-availability}

To ensure every Lyquid service has a fair and smooth start, Lyquor introduces \emph{Archival Nodes}, a decentralized set of nodes that replicate the network state for all Lyquids. These nodes provide \emph{baseline on-chain data availability} and guaranteed onboarding for new services.

Archival nodes run the same Lyquor program but operate in an inclusive mode. They are \emph{not} required to host the off-chain part of each Lyquid, which makes them lightweight and broadly compatible. They also do not need per-Lyquid specialized hardware or capabilities required by off-chain computation.

Importantly, archival nodes form only a \emph{small subset} of the network to bootstrap a Lyquid. The majority of Lyquor nodes participate selectively, hosting only the Lyquids they choose based on utility, performance, or economic incentives. This forms a market-driven execution landscape where hosting naturally concentrates around useful services, and underutilized ones fade without needing protocol-level intervention.

While archival nodes ensure onboarding and state completeness, they also play a critical role in supporting FCO (Section~\ref{sec:fco}). By maintaining access to global side effects, they allow composable contract behavior even when execution is distributed across selectively hosted nodes.

\subsection{Fearless Off-chain, Serverless Computation}
\label{sec:offchain-computation}
While most platforms treat off-chain logic as an afterthought, Lyquor makes it a first-class citizen. In traditional systems, off-chain tasks often rely on external availability layers, intermediate indexing services, or centralized infrastructure. This introduces latency, increases system complexity, and creates new trust boundaries. Worse, off-chain code typically loses the simplicity that makes smart contracts appealing in the first place. Developers value smart contracts because they are like FaaS~\cite{faas}, but without hassle in data preservation. That same convenience has rarely been available for fast, off-chain workloads.

Lyquor changes that. With Direct Memory Architecture (DMA) (Section~\ref{sec:dma}), off-chain logic runs against the same structured state as on-chain execution: no schema design, no database plumbing. Developers simply write stateful logic, without infrastructure baggage. This combines the best of both worlds: the low-friction experience of smart contracts with the speed and flexibility of serverless functions~\cite{serverless}.

While selective hosting in Lyquor elegantly resolves the longstanding tradeoff between scalability and composability faced by traditional Layer-1/2 platforms, its original motivation was even deeper: enabling \emph{seamless and resilient off-chain computation}. Off-chain workloads vary widely in hardware requirements and performance characteristics, and differ fundamentally from on-chain computation where nodes must replicate the \emph{same} execution. In contrast, off-chain tasks often involve different nodes performing \emph{different} portions of a larger computation by specialization.

Without selective hosting, every node would be forced to carry the on-chain state of every contract, including those irrelevant to its off-chain responsibilities. This creates unnecessary deployment barriers and breaks the seamless flow between on-chain and off-chain logic. As a result, most existing off-chain systems rely on trusted Data Availability layers to fetch inputs. The actual computation is often performed by centralized, one-off implementations, typically written without a common framework. Unlike on-chain contracts, these off-chain components lack standardization, code validation, and result resilience, which makes them difficult to trust, scale, or generalize across applications.

In addition to accessing on-chain data, off-chain logic in Lyquor can also collaborate across nodes in a \emph{programmable} way. This is enabled by Lyquor's Universal Procedure Call (UPC) (Section~\ref{sec:upc}), a multicast function call that allows a Lyquid service to invoke custom logic on multiple nodes concurrently. Each node can perform part of the task or act in specialized roles. Results are then aggregated or reconciled by the caller in real time. This model supports resilient coordination across untrusted infrastructure without sacrificing performance.

\subsection{Lightweight Execution, Cloud-Like Economics}
\label{sec:cloud-fee-model}

Because Lyquid execution is powered by DMA (Section~\ref{sec:dma}), on-chain calls are fast, lightweight, and direct: with no serialization, no object decoding, and no key hashing. The result is a snappier execution layer with minimal runtime overhead.

Crucially, this means \emph{on-chain execution now costs as little as off-chain logic}, closing the historical gap that made blockchain compute expensive and over-metered. In Lyquid, both on-chain and off-chain logic share the same execution engine and data model. The only distinction is that on-chain execution is triggered by events sequenced on the chain. With this symmetry, it becomes possible to rethink not just performance, but also how computation is priced.

Instead of relying on fine-grained gas metering, Lyquor could adopt a simpler model:

\begin{itemize}
    \item Each call is bounded by a maximum gas limit (set generously, thanks to DMA's efficiency).
    \item Actual fees could be application-defined. Developers may implement their own pricing logic inside their Lyquid.
\end{itemize}

This mirrors the economics of cloud services, where users are billed per API call rather than per CPU instruction. Nodes, in turn, decide whether a service is worth running based on real economic value, not just technical cost.

While this fee model remains open to future refinement, it introduces exciting possibilities:

\begin{itemize}
    \item No overpricing for simple logic.
    \item No forced duplication of execution.
    \item Less protocol-imposed friction, more developer-driven pricing flexibility.
\end{itemize}

As the Lyquor ecosystem matures, this model can evolve naturally, blending \emph{performance efficiency with economic simplicity}, and giving developers more control over how their services are accessed and monetized.

To summarize, Lyquor isn't a Layer-1 or a Layer-2. It's a platform rethinking the blockchain infrastructure model entirely:

\begin{itemize}
    \item Nodes run services, not chains.
    \item Execution is modular and selective.
    \item Composability is preserved via global sequencing.
    \item Off-chain logic becomes the first-class citizen with the same model.
    \item Developers control economics, not just logic.
    \item Scalability emerges organically, service by service.
\end{itemize}
This is not just a new contract framework. It's what a \emph{cloud-native, service-aligned decentralized} compute network should look like.

This section presented Lyquor's core concepts and architectural vision. The remainder of the paper examines each technical component in detail: Lyquid programming model, Fate-Constrained Ordering, Direct Memory Architecture, and Universal Procedure Call.

\section{Why Lyquid?}
\label{sec:lyquid}

Lyquid represents a paradigm shift in blockchain development by offering several key advantages.

\paragraph{Unified Programming Model}
Traditional blockchain development forces you to separate your on-chain and off-chain logic, often requiring different languages, frameworks, and coordination layers. Lyquid simplifies this by unifying both under a single programming model, where \emph{on-chain and off-chain state coexist seamlessly}, and logic is written side by side.

\begin{minted}{rust}
lyquid::state! {
    // On-chain state (like Solidity storage)
    network total_supply: U256 = U256::ZERO;
    network balances: network::HashMap<Address, U256> = network::new_hashmap();
    
    // Off-chain state (unique to Lyquid)
    instance local_transactions: Vec<Transaction> = Vec::new();
}
\end{minted}

With Lyquid, your off-chain computation code is bundled together with the on-chain logic in a hash-verified package. This ensures that Lyquor nodes, which execute the on-chain logic faithfully, also run the corresponding off-chain logic consistently, eliminating the need for external or bespoke off-chain infrastructure.

However, because this logic runs off-chain, it can still react independently to external events and access each node's own off-chain, instance-local state. This makes it possible to build efficient distributed and decentralized protocols, where all nodes execute consistent logic, while still interacting with their own unique local environment, which unlocks new classes of applications.

\paragraph{Ethereum Friendliness and API Compatibility}
\label{sec:eth-friendly}

If you're coming from Solidity, you'll feel right at home. Lyquid's programming model aligns naturally with the mental model of EVM developers, but with all the benefits of modern Rust semantics and Lyquor's powerful execution model.

Here's a quick comparison using a familiar example (an ERC-20 \texttt{transfer} function):

\begin{minted}{rust}
// Solidity
contract ERC20 {
    mapping(address => uint256) private _balances;
    /* ... other state variables ... */
    
    function transfer(address to, uint256 amount) public returns (bool) {
        address owner = msg.sender;
        _transfer(owner, to, amount);
        return true;
    }
    /* ... other functions ... */
}
\end{minted}

\begin{minted}{rust}
// Lyquid
lyquid::state! {
    network balances: network::HashMap<Address, U256> = network::new_hashmap();
    /* ... other state variables ... */
}

lyquid::method! {
    network fn transfer(&mut ctx, to: Address, amount: U256) -> LyquidResult<bool> {
        let from = ctx.caller.clone();  // equivalent to msg.sender
        transfer(&mut ctx.network, from, to, amount)?;
        Ok(true)
    }
    /* ... other functions ... */
}
\end{minted}

The parallels between Solidity and Lyquid are intuitive:
\begin{itemize}
    \item Solidity's mapping $\Leftrightarrow$ Lyquid's network::HashMap
    \item Solidity's msg.sender $\Leftrightarrow$ Lyquid's ctx.caller
    \item Solidity's public functions $\Leftrightarrow$ Lyquid's network functions (\texttt{network fn ...})
\end{itemize}

In addition, during the Lyquid build process, our compiler tools automatically generate compatibility layers that expose your contract's methods via standard Ethereum APIs like \texttt{eth\_call} and \texttt{eth\_sendTransaction}. As long as your function signatures use supported types (e.g., integers, strings, vectors, fixed arrays, etc.), your contracts can be called directly through Ethereum wallets. No custom SDKs or wrappers are required.

This means you can build next-generation contracts with Lyquid while still using familiar wallets, RPC tooling, and DApp interfaces, combining forward-looking capabilities with backward-compatible ergonomics.

\paragraph{Invisible Power, Built Right In}
Lyquid doesn't just simplify development. It embeds powerful features that redefine smart contract execution. Our Direct Memory Architecture (DMA) (Section~\ref{sec:dma}) eliminates the gap between on-chain and off-chain computation by enabling on-chain logic and state to run as efficiently as off-chain, container-style code. This approach erases the traditional performance disparities, delivering rapid, cost-effective processing without sacrificing security. Conversely, DMA lets off-chain computation enjoy the same simplicity and programmability as on-chain logic, offering a more intuitive, stateful programming model compared to typical function-style serverless solutions.

Complementing DMA, the Universal Procedure Call (UPC) (Section~\ref{sec:upc}) offers a fault-tolerant, decentralized primitive for off-chain computation. Unlike many current off-chain solutions which rely on ad hoc, centralized setups that sacrifice decentralization and reliability, UPC provides a concise, resilient pattern to coordinate off-chain execution across the network. This approach simplifies the development of complex distributed protocols while restoring the benefits of true decentralization for off-chain operations.

Together, DMA and UPC empower Lyquid to deliver a next-generation smart contract/serverless computing platform that offers high-performance execution and robust, decentralized off-chain coordination, all within a unified programming model.

{\centering
\footnotesize
\renewcommand{\arraystretch}{1.25} 
\begin{tabular}{|l|p{3cm}|p{4cm}|p{4.5cm}|}
\hline
\rowcolor{gray!20} 
\textbf{Feature} & \textbf{Solidity/EVM} & \textbf{Rust/WASM}\newline\textbf{(Solana, ink!, CosmWasm, ICP)} & \textbf{Lyquid} \\
\hline
\textbf{On-chain Persistent State} & \cmark{} (\texttt{storage}, slots) & \cmark{} (serialized key-value objects) & \cmark{} (\texttt{network} state) \\
\hline
\textbf{Off-chain Local State} & \xmark{} Not supported & \xmark{} Requires external indexing or workers & \cmark{} Native persistent (\texttt{instance} state) \\
\hline
\textbf{Native Off-chain Computation} & \xmark{} Not supported & \xmark{} External solutions required & \cmark{} On-chain style, per-node behavior with verified code \\
\hline
\textbf{Developer Ergonomics} & \ymark{} Custom DSL, limited types & \ymark{} Rust-native but complex setup, explicit serialization & \cmark{} Rust-native, minimal macros, easy setup \\
\hline
\textbf{Execution Efficiency} & \xmark{} Slow, high gas overhead & \ymark{} Faster instruction speed, more state overhead & \cmark{} Consistent high efficiency on-chain \& off-chain (DMA, minimal overhead) \\
\hline
\end{tabular}
\captionof{table}{Comparison of different VMs (and their DSLs).}
}

\section{Fate-Constrained Ordering}
\label{sec:fco}

The key change we made by switching from a Chain-Centric to a Service-Centric design, compared to all other platforms, is to flip around two fundamental assumptions of blockchain architecture.
\subsection{Top-Down vs. Bottom-Up}
Most blockchain platforms today adopt a \emph{bottom-up} approach to scalability. They establish \emph{local} sequences (ledgers, chains, or state machines) within groups (committees) of nodes, followed by a \emph{global} stage that \emph{settles} interactions across these groups. This pattern is common in systems like Avalanche's subnets, the Cosmos Network, Polkadot parachains, Near's sharded execution, and Ethereum Layer-2 platforms such as Optimism~\cite{optimism} and Arbitrum~\cite{arbitrum}.

\begin{figure}[h]
\centering
    \includegraphics[width=0.6\textwidth]{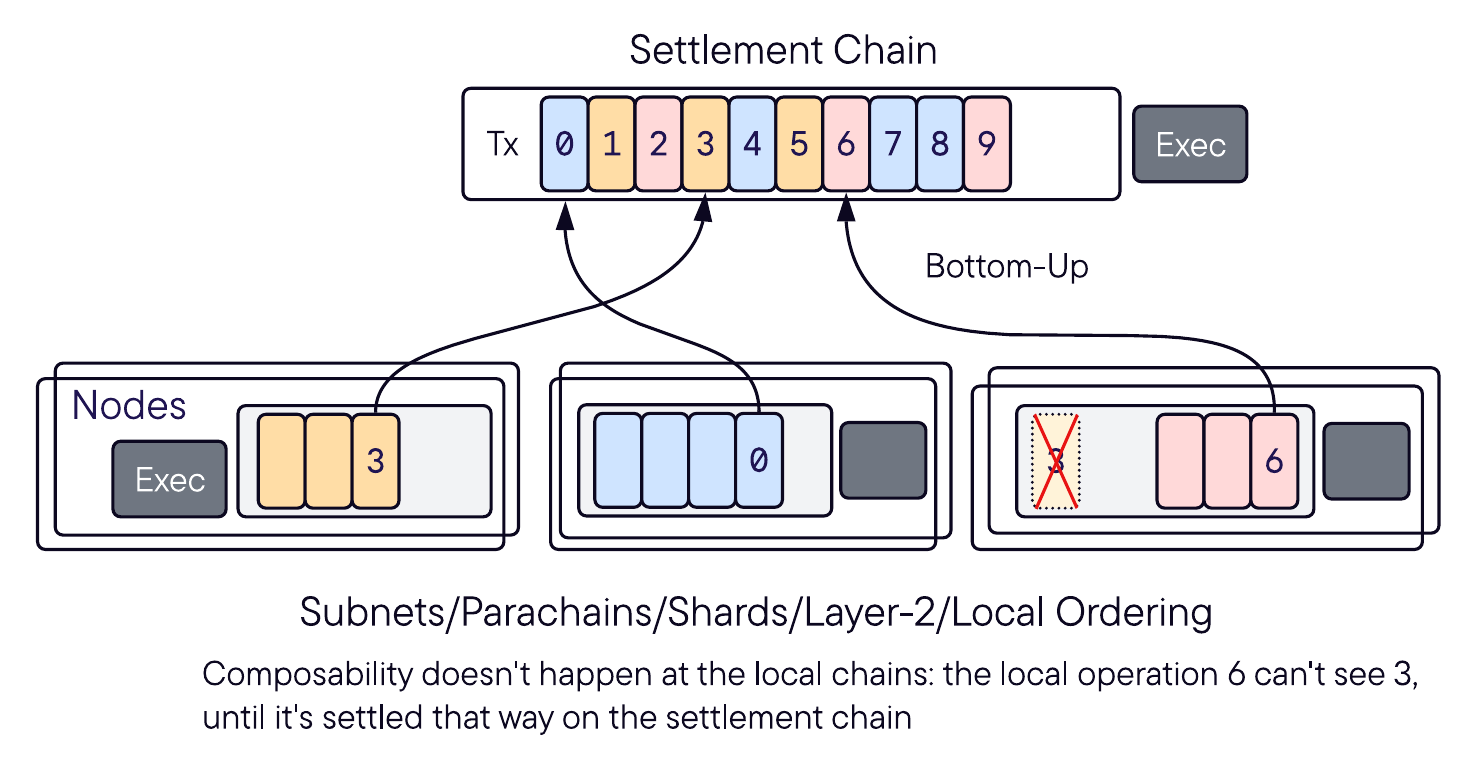}
    \caption{Bottom-up scalability approach: local sequences within node groups followed by global settlement.}
    \label{fig:fco-bottom-up}
\end{figure}

The rationale of partitioning nodes into several groups stems from the nature that many contracts are irrelevant or only need inter-operations occasionally. The contracts can be categorized into these groups so they get more dedicated handling for better performance and waste fewer resources. This resembles a bank having branches in different cities as an amenity to the locals and offload the processing burden by geographic distribution of their customers.

This approach enables scalability through validator specialization but introduces several architectural problems:

\paragraph{Ordering Uncertainty} Because each group has its own local consensus, the order of transactions is only finalized within that group. Interactions that span multiple groups are not globally ordered at the time they occur. The global settlement layer must reconcile these interactions after the fact, which opens the door to race conditions and unpredictable outcomes.

\paragraph{Loss of Atomicity} Unlike centralized, monolithic systems like banks, in a decentralized deployment, once transactions cross group boundaries, it becomes difficult to guarantee all-or-nothing behavior across multiple contracts. The lack of a shared execution context breaks end-to-end transactional guarantees.

Some platforms attempt to avoid fragmentation altogether. Solana~\cite{solana}, for example, uses a monolithic architecture where all contracts are executed by a single global validator set. While this preserves composability, it requires full-state replication and limits scalability through specialization. Others, like Sui~\cite{sui}, adopt an object-centric model that allows parallel execution when dependencies are absent, but still requires a second coordination phase when causal relationships arise. This effectively results in a two-stage composability model, which behaves similarly to bottom-up systems under contention.

\begin{figure}[h]
\centering
    \includegraphics[width=0.6\textwidth]{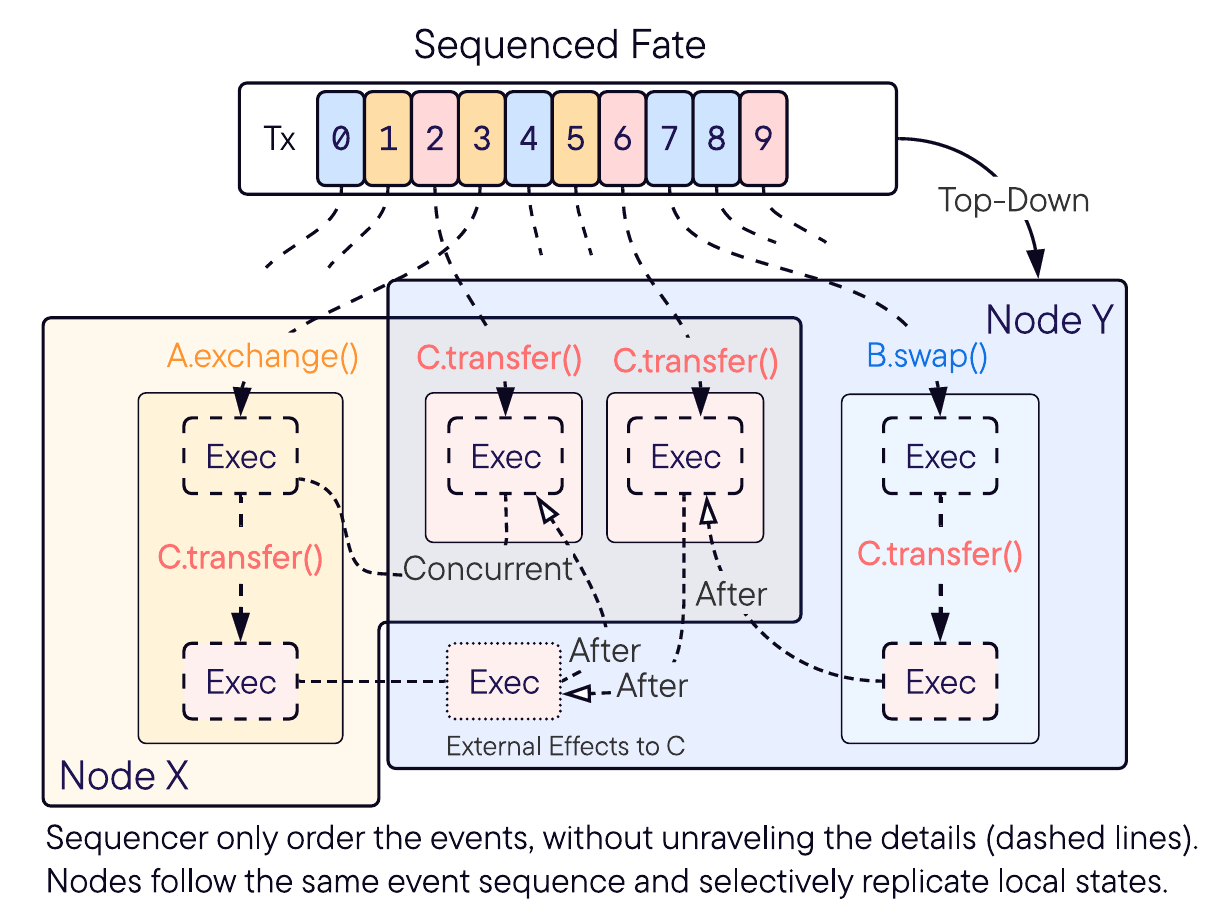}
    \caption{Lyquor's top-down approach: global sequencing first, then distributed execution by specialized nodes.}
    \label{fig:fco-top-down}
\end{figure}

In Lyquor, we address these challenges by flipping the model around. Instead of scaling from local ordering upward, we start with a global sequence:

\begin{enumerate}
    \item We first decouple sequencing (consensus) from execution (VM), enabling a lightweight consensus layer to determine the global, total order of all contract interactions. This ordering is established efficiently, without requiring any node to interpret or execute the transactions during consensus.
    \item Lyquor moves from a Chain-Centric architecture to Service-Centric. Execution happens per service by a \emph{contract-like} unit named Lyquid (Section~\ref{sec:lyquid}). Each Lyquid can be hosted by a group of nodes. This means nodes are no longer grouped by chains anymore like the bottom-up approach, but through the hosting of Lyquids. This offers way more flexibility that nodes can then organically select Lyquids to host, and can freely participate in multiple groups simultaneously to max out their computation capabilities with specialization for their best profits.
    \item As the sequence is finalized, each node independently follows the global order and executes the subsequence for each Lyquid it has chosen to host. A node can host all Lyquids and execute the full sequence, or it can selectively run only the Lyquids that balance its incentives or capabilities.
    \item Importantly, executing Lyquids does not require the node to participate in consensus. While a node may be configured to handle both roles, Lyquor treats consensus voting and execution as distinct responsibilities, and expects them to be carried out by separate networks.
    \item Even if a node executes only a subsequence corresponding to a specific Lyquid (or a set of Lyquids), it is still aware of the effects by the total order of Lyquids it did not run. The fate of the system is fixed through the global order, and each node simply unravels the portion it touches, preserving composability without requiring full execution.
\end{enumerate}

This notion of fate sharply contrasts with the behavior of bottom-up systems. In those architectures, global outcomes emerge only after separate node groups finalize their local results and attempt to reconcile them. The actual order of effects across contracts in different groups is not known until global settlement, and after other factors like bridge logic, or external relayers. In Lyquor, the fate of the system (the timeline of all Lyquid calls) is determined at the global sequencing stage, before any execution begins, and must be respected by all nodes.

This is similar to a play script that has already been written. Each node picks up the script and performs only the scenes relevant to its role. The storyline is not improvised or invented during performance. It is already there, waiting to be revealed. Execution does not define what happens; it discovers what was already determined. This is the essence of Fate-Constrained Ordering (FCO): a global ordering that allows every node to reason about Lyquid interactions reliably, even while executing only a small slice of the system.

This design principle coincides with the ideas explored in Corfu~\cite{corfu}, a foundational system for cloud-scale replication. Lyquor, like Corfu, separates the act of ordering operations from the act of executing state, using a shared log to impose a total order while allowing each replica to materialize state on demand. Drawing from the world of distributed systems, Lyquor carries forward the philosophy behind large-scale, efficient infrastructure and applies it to decentralized environments.

\subsection{Full Nodes vs. Light Nodes}

Traditional L1 platforms distinguish sharply between full and light nodes. Full nodes are responsible for both participating in consensus and executing every transaction and maintain hosting capabilities for all contracts. They are expected to act as consistent replicas of the system, maintaining identical state transitions and outputs.

Over time, the entry barrier to becoming a full node has grown significantly. This is a natural consequence of expecting every node to replicate all contracts, both in computation and storage. As the diversity and complexity of deployed applications increases, so does the cost of replication. In response, many platforms introduced the notion of light nodes, which do not vote in consensus or replicate contract state locally. They instead keep proofs or query full nodes to track specific states, but do not add to the security of the core network.

In Lyquor, we flip the assumptions around in three changes:

\paragraph{Lightweight, Independent Nodes} We enrich the capabilities of a light node, so it effectively becomes a validator that locally replicates the selected contracts' state by following the sequencing result. Each Lyquid may contain both network state, shared by hosting group of nodes, driven by sequencing; and instance state, local to each node in the group, driven by off-sequence events. Nodes participate directly in service provisioning as a server or get embedded directly into wallets or other user-end apps as a client.

\paragraph{Slim, Global Composability Archive} We trim down what full nodes should do in the networks and refer to them as \emph{Archival Nodes} in Lyquor. They are \emph{not required} to render services to users, but to execute the globally ordered sequence of \emph{all} Lyquids, for two main purposes:
\begin{enumerate}
    \item Supplying side effects and output data for contracts not hosted by a given node, and
    \item Ensuring base availability (Section~\ref{sec:base-availability}) for any Lyquid, especially during its onboarding phase.
\end{enumerate}
Although archival nodes may still behave like regular nodes if desired, their essential role lies in composability and availability, not service rendering.

\paragraph{Service-First Scalability} The majority of the network is made of lightweight regular nodes, not the archival nodes. Once quickly bootstrapped, a regular node can just use its local information for all services.

Additionally, archival nodes can offer hints for performance optimization. When a regular node executes the subsequences, archival nodes may help the regular node skip redundant work or understand runtime dependencies better. These hints do not affect correctness or determinism. They are simply early exposures of an already-fixed ordering, revealed for efficiency.

This model requires no separate software stacks. Archival and regular nodes run the same protocol but follow different strategies. Archival nodes are general-purpose: they require no specialized hardware and earn stable, universal staking rewards for maintaining full availability. Regular nodes, in contrast, curate a portfolio of profitable staking of contracts. They can earn more, but their returns depend on demand and may require specialized resources, driven by the services they choose to host.

To summarize, partial execution is the new norm, and full-state retention is secondary. Rather than layering light nodes as an afterthought with centralized API which undermines security, Lyquor makes decentralization and specialization a coherent starting point. Scalability emerges naturally from a model where nodes are free to choose their scope of execution, without sacrificing global composability.

\subsection{Example: Two DEX Contracts}
To better understand how Fate-Constrained Ordering (FCO) works in practice, let’s consider an example involving two DEX contracts, A and B. For simplicity, we'll focus on a shared liquidity token C, which both DEXes use for asset exchange.

Assume there is a node X that is only interested in DEX A. It does not wish to host DEX B. However, since both A and B involve contract C to move tokens, any side effect B causes to C, such as modifying a user's balance, must still be visible from A’s point of view, even though X does not care about B directly.

\begin{figure}[h]
\centering
    \includegraphics[width=0.6\textwidth]{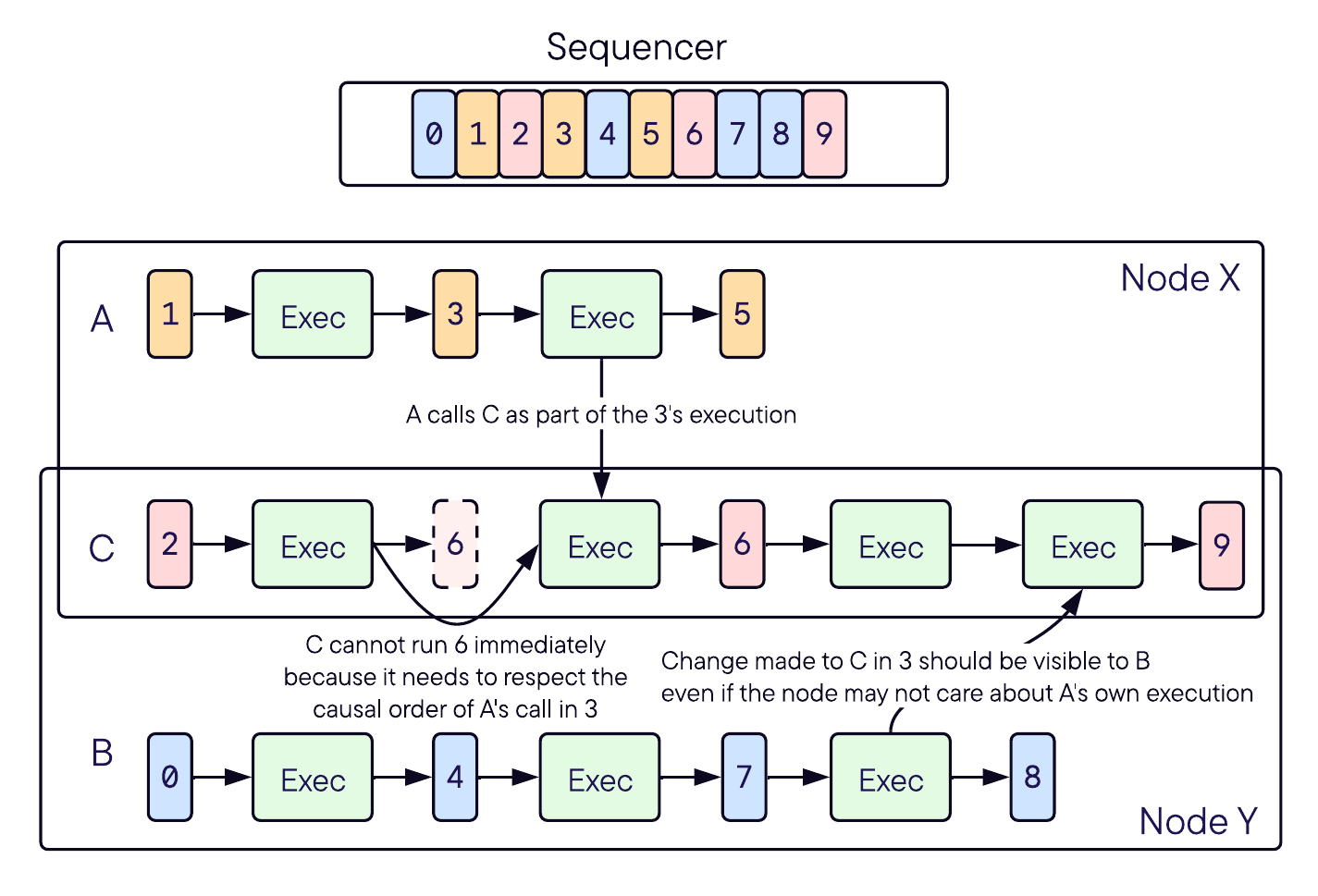}
    \caption{FCO example: Two DEX contracts (A and B) sharing liquidity token C, showing how partial execution maintains global consistency.}
    \label{fig:fco-example}
\end{figure}

In this example, FCO guarantees composability while allowing partial execution. Let’s walk through the key steps:

\begin{enumerate}
    \item Node X hosts A and C, since C is invoked by A. While it is possible to skip executing C and retrieve all A-to-C call results from archival nodes, executing C locally is often preferred for efficiency and trust minimization, especially when C is a widely used token like USDC.
    \item Meanwhile, another node Y is only interested in DEX B, so it hosts B and C. Since C is a shared liquidity contract, its state must remain consistent across both A and B. For example, a user's balance in C must reflect all relevant transactions, regardless of which contract caused the change. This ensures no double spend or state inconsistency.
    \item When node X executes transaction 3, which calls a function in contract A, that function may internally invoke a function in C. To preserve global consistency, node X must schedule transaction 6, which modifies C, after transaction 3 according to the global sequence. For example, transaction 3 may already drain the user's remaining balance in C. If transaction 6 then attempts to transfer additional tokens, it should fail. If C were to process its transactions independently without respecting this order, the resulting state would be incorrect.
    \item Similarly, when node Y executes transaction 7, which invokes C from contract B, the effects of A's earlier execution (up to transaction 7) must already be visible. Even though node Y did not run A, the changes made to C are part of the global fate and must be respected. These external effects can be learned from archival nodes.
\end{enumerate}

If there is no causal interaction at a given point in the sequence, for example when A and B are not currently calling the shared contract C, they can still be executed in parallel. This holds even if both contracts depend on C elsewhere in the sequence. FCO enables this fine-grained concurrency by enforcing only the ordering constraints that actually exist at each step, allowing unrelated parts of the execution to proceed independently while preserving global consistency.

It is also worth noting that even though node X must observe all B-to-C interactions to ensure consistency, it does not need to understand how B arrived at those calls. X only needs to observe the terminating effects on C, not the internal execution path inside B or any of B’s interactions with other tokens like D. This property allows nodes to remain lightweight and focused, even when composability spans across multiple contracts.

This example illustrates how FCO enables seamless composability across selectively hosted contracts. Nodes can independently execute only what they care about, while still respecting the global ordering of the system.

\section{Direct Memory Architecture}
\label{sec:dma}

Although many smart contract platforms today advertise "native-speed execution", especially those built with high-performance WASM runtimes, this often overlooks the real bottleneck in blockchain computation. In practice, most smart contract transactions are extremely \emph{short-lived}, performing only a few arithmetic or conditional operations. The real cost lies in how state is accessed and persisted.

Take a typical DEX swap operation as an example: while the core logic involves just a few integer calculations to apply a constant product formula ($x \cdot y = k$), the surrounding state access is far more expensive. Reading and updating balances and liquidity pool reserves often requires multiple storage lookups, key hashing, and object (de)serialization, all of which introduce overhead that \emph{far exceeds the actual computation}.

Lyquor addresses this fundamental bottleneck with a radically different approach: each contract is given its own virtual, non-volatile, byte-addressable memory space, similar to the memory model of a cloud virtual instance that can be suspended, rather than a database. This memory is persistent across executions, yet fully portable and virtualized, requiring no special hardware or OS modifications. State access becomes as efficient and intuitive as working with in-memory data structures, eliminating the need for serialization, hashing, or storage layout gymnastics.

\subsection{Benefits}
\paragraph{Zero Translation} There is only one representation: in-memory form. No conversion overhead.
\paragraph{On-Demand Loading} Only the needed fraction of the virtual memory is loaded when accessed.
\paragraph{Efficient Caching} Frequently used portions of memory (e.g., accounts, protocol states) stay hot in host's OS, enabling native performance.
\paragraph{Unified Access} Both network and instance states are handled through the same memory with different address range. Off-chain code can manipulate instance state that is persisted per node. Network state memory is correctly versioned during the execution of the sequence from FCO (Section~\ref{sec:fco}).
\paragraph{Data-structure Agnostic} Developers can use third-party Rust libraries: as long as they support a custom \texttt{Allocator}~\cite{allocator}, already provided in Lyquid LDK. Nothing else needs to change. To simplify development, the LDK also ships with common standard containers (\texttt{Vec}, \texttt{HashMap}, etc.) under \texttt{runtime::network}~\cite{runtime-network} and \texttt{runtime::instance}~\cite{runtime-instance}.
\paragraph{Portable Node} DMA is implemented entirely in \emph{user space}, and therefore \emph{does not require} special OS/kernel tweaks. Our node can run on major Unix-like platforms and architectures such as macOS/Linux/WSL and x86/arm. It can run on Android without rooting the device.

\subsection{The Real Bottleneck in State Access}

From years of our research and industrial experience, state access overhead is one of the \emph{most underestimated} performance bottlenecks in blockchain systems, regardless of whether it's L1 or L2, sequential or parallel execution.

Even in those ultra-fast, consensus-free L2 systems, all transactions must eventually access the state (at least once, for ZK-rollups). In parallel execution environments, contention still occurs when multiple transactions access the same contract, often rendering parallelism ineffective due to shared-state synchronization.

There have been many different “Virtual Machines” (VMs) used across L1 and L2 blockchains. While they differ in language specification and execution performance, they all share a common goal: to make smart contract development easier by \emph{abstracting away low-level storage and database details}. Broadly, there are two dominant approaches:

\paragraph{Slot-based} This approach is exemplified by the Ethereum Virtual Machine (EVM). In Solidity, the most widely used language compiled to EVM bytecode, the compiler must translate high-level data structures like bytes, arrays, and mappings into a restricted word-level key-value mapping model, because EVM provides only two core operations for persisted state access:
\begin{itemize}
    \item \texttt{sload}: read a 32-byte word from a key.
    \item \texttt{sstore}: store a 32-byte word to a key.
\end{itemize}
Solidity partitions all variables into logical "slots"~\cite{slots} starting from 0 index, where each one stores a 32-byte data chunk. The identifiers for the slots are then hashed into fixed 32-byte keys, which the EVM uses for load/store instructions.

\paragraph{Object-based} This approach is more coarse-grained. Platforms like Solana~\cite{solana}/Polkadot~\cite{polkadot}/Dfinity~\cite{dfinity} treat on-chain state as entire serialized objects (of structs). These objects are defined by the user and serialized/deserialized either manually or via automatic derivation (e.g., Rust's serde~\cite{serde}), typically stored under hashed keys (e.g., account-derived addresses or internal storage hashing). State updates occur when transactions invoke functions that mutate the in-memory representation, which is later serialized back.

{
\centering
    \includegraphics[width=0.7\textwidth]{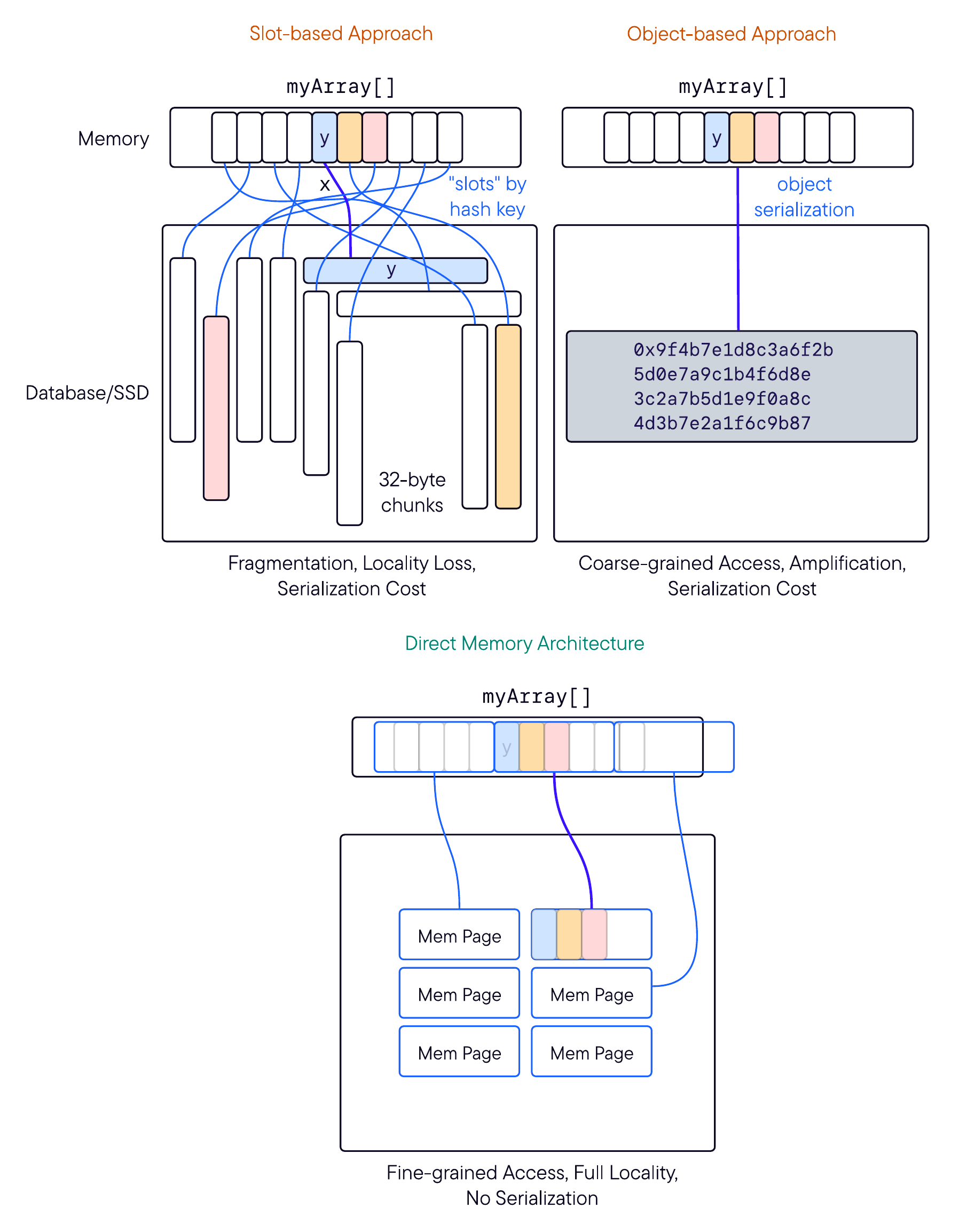}
    \captionof{figure}{Comparison of slot-based, object-based, and DMA approaches to state management in smart contracts.}
    \label{fig:dma}
}

\subsection{Why DMA is Better?}

There are pros and cons for each of the two approaches:

\paragraph{Slot-based}
\begin{itemize}
    \item Allows partial state access, which reduces amplification (i.e., how much extra data is accessed beyond what's precisely needed).
    \item Offers reasonable developer ergonomics, as Solidity hides the complexity of storage translation.
    \item However, hashing each slot destroys data locality. For example, an array of uint64 values becomes fragmented into isolated slots with "random" hashed keys.
    \item Developers are largely confined to using only two kinds of data structures offered by Solidity: mappings and arrays, as the compiler applies special logic to layout these structures somewhat efficiently across storage slots. Technically, more complex structs are possible, but require custom layout handling and are practically discouraged.
\end{itemize}

\paragraph{Object-based}
\begin{itemize}
    \item Developers face challenges in designing efficient serialization/deserialization logic.
    \item Although deriving Serialize/Deserialize in Rust seems straightforward in toy examples, it's not scalable for large, realistic states (e.g., user balances, liquidity pools, order books, multi-megabyte state).
    \item Platforms try to provide optimized containers (like \texttt{Vec}, \texttt{BTreeMap}, \texttt{HashMap}), but this leads to several limitations:
    \begin{enumerate}
        \item Platforms must optimize and customize these containers internally, often using smart pointers or internal metadata tracking to capture writes made at finer granularity, effectively reintroducing slot-based storage mechanics under the hood.
        \item Users are restricted to these "blessed" containers. Using third-party data structures requires deep modifications (e.g., custom pointers), which is impractical.
        \item Even then, platforms can't introspect or incrementally access \emph{user-defined types inside these containers}, so they still require full serialization and deserialization for each contract execution, again losing fine-grained access capabilities.
    \end{enumerate}
\end{itemize}

Lyquor addresses these challenges by treating each contract hosted on a node as a lightweight container with proper memory isolation and dedicated, persistent (non-volatile) memory. The network state of each contract is driven by sequences from FCO, while its instance state is shaped by other per-node events.

\begin{table}[h!]
\centering
\footnotesize
\renewcommand{\arraystretch}{1.25} 
\begin{tabular}{|l|c|c|p{6cm}|}
\hline
\rowcolor{gray!20} 
\textbf{Feature} & \textbf{Slot-based} & \textbf{Object-based} & \textbf{DMA} \\
\hline
\textbf{Fine-grained Access} & \cmark{} & \xmark{} * & \cmark{} (byte-level) \\
\hline
\textbf{Cache Locality} & \xmark{} & \ymark{} & \cmark{} (dynamic page caching) \\
\hline
\textbf{Low Fragmentation} & \xmark{} & \cmark{} & \cmark{} \\
\hline
\textbf{No Manual Layout} & \cmark{} & \xmark{} * & \cmark{} \\
\hline
\textbf{Good Ergonomics} & \ymark{} & \xmark{} * & \cmark{} (in-memory allocation) \\
\hline
\textbf{Rich Data Structures} & \xmark{} * & \ymark{} & \cmark{} (\textit{any} data structure via \texttt{Allocator}) \\
\hline
\textbf{Hash-free} & \xmark{} & \xmark{} & \cmark{} (32-bit memory address) \\
\hline
\end{tabular}
\caption{Comparison of memory management models.}
\label{tab:memory_models}
\end{table}

\section{Universal Procedure Call}
\label{sec:upc}

Lyquid introduces Universal Procedure Call (UPC), a powerful built-in primitive that extends the capabilities of smart contracts by enabling \emph{seamless and composable off-chain communication} across nodes and services. UPC combines the simplicity of local function calls with the power of multicast, fault-tolerant, distributed execution, making it a foundational building block for advanced decentralized applications.

\subsection{What is UPC?}
At its core, UPC allows a Lyquid to invoke off-chain logic across multiple nodes. Each invocation uses the same input, but nodes may execute the procedure using their own \emph{node-local state}, making UPC useful not just for redundancy, but also for distributed computation and coordination.

Unlike traditional RPC systems, UPC is \emph{deeply integrated into Lyquid's programming model}. It is expressed as a normal Lyquid function: no boilerplate, no special networking setup. Developers can write UPC handlers and initiate UPC calls from standard instance functions, which are \emph{fully compatible} with Ethereum APIs. This means wallets can transparently benefit from decentralized request handling and fault-tolerant execution without any frontend changes.

\begin{figure}[h]
\centering
    \includegraphics[width=0.7\textwidth]{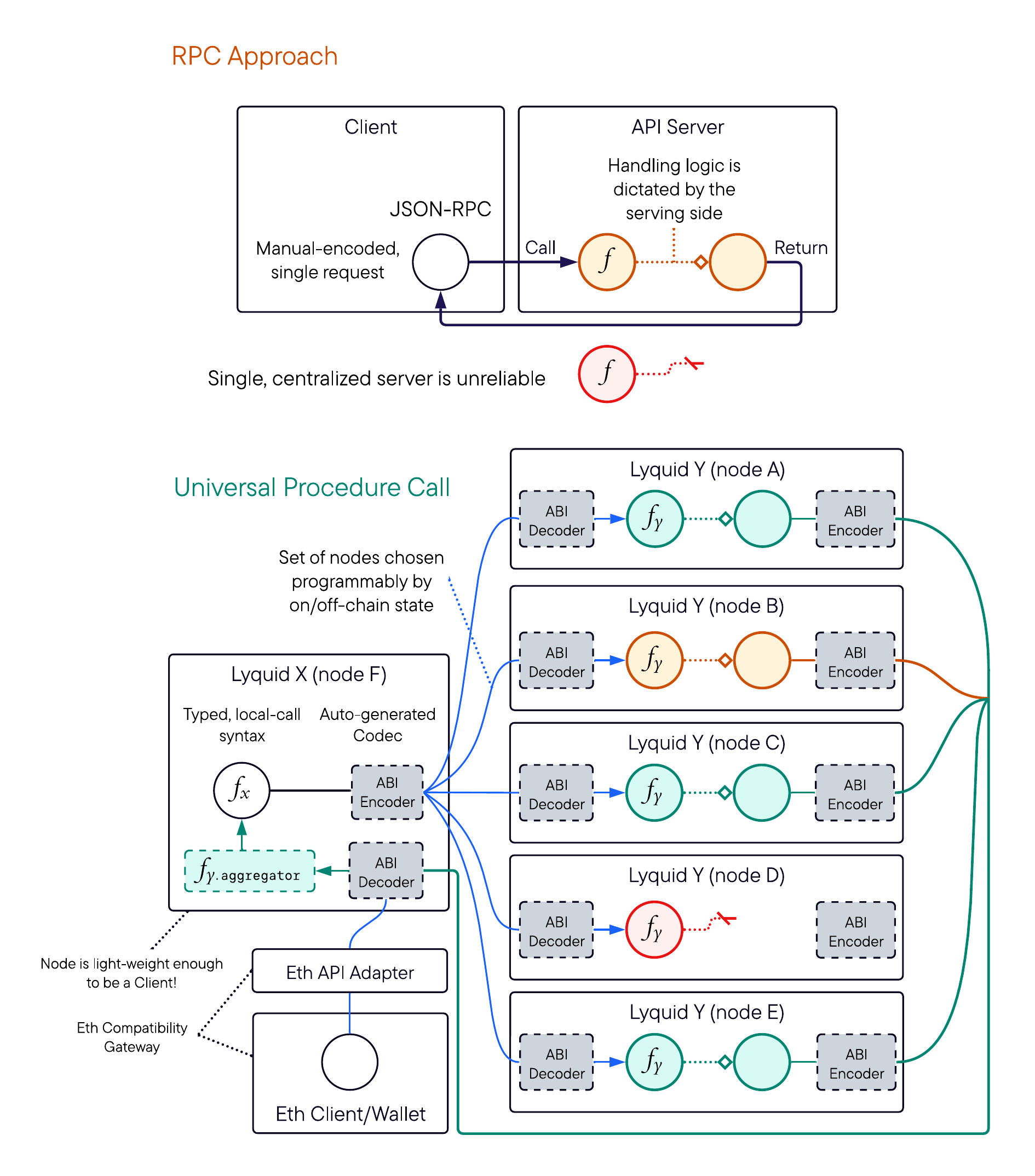}
    \caption{Multicast function calls enabling distributed computation and fault-tolerant coordination across multiple nodes.}
    \label{fig:upc}
\end{figure}

\subsection{Typical Styles}
There are some typical execution and aggregation styles that one can easily express with UPC supports, depending on the application's needs:

{
\centering
\footnotesize
\renewcommand{\arraystretch}{1.25}
\begin{tabular}{|>{\raggedright}p{4cm}|>{\raggedright}p{4cm}|>{\raggedright}p{3.5cm}|>{\raggedright\arraybackslash}p{4cm}|}
\hline
\rowcolor{gray!20}
\textbf{Mode / Pattern} & \textbf{Per-Node Behavior} & \textbf{Aggregation Strategy} & \textbf{Common Use Cases} \\
\hline
\textbf{1. Share-based Aggregation} & Each node returns a unique share or partial result & Aggregate shares into a final result (e.g., via threshold logic) & Oracle and Bridge serving, Threshold signatures, Verifiable Random Functions (VRF), Multi-Party Computation (MPC) \\
\hline
\textbf{2. Availability} & All nodes execute with the same input, and expect the same output & Accept first valid response (e.g., hash match, on-chain verification) & High-performance decentralized storage, Resilient data transmission, Wallet serving \\
\hline
\textbf{3. Distributed Computation} & Nodes return distinct partial results (state-dependent) & Custom, step-by-step logic & Decentralized orderbook matching, AI workload orchestration, Reputation scoring, Cloud-style service mesh coordination \\
\hline
\textbf{4. Simple RPC} & One selected node executes the UPC handler & Return the result directly & Lightweight service calls, internal routing logic \\
\hline
\end{tabular}
\captionof{table}{Comparison of distributed computation patterns.}
\label{tab:computation_patterns}
}

\subsection{More Than Just Multicast}
While multicast actor models have existed in distributed systems and game engines (e.g., Unreal Engine~\cite{unrealengine}), UPC goes far beyond by making the \emph{coordination layer itself programmable and state-aware}. What makes UPC fundamentally more powerful is how it is deeply integrated into Lyquid's execution and state model, not just bolted on as an extension.

\paragraph{Programmable Node Selection}
The decision of which nodes should execute a UPC handler is not hardcoded. It can be \emph{contract-defined}, driven by:
\begin{itemize}
    \item \emph{Network state}, such as the virtual cluster topology, validator reputation scores, availability metrics.
    \item \emph{Instance state}, such as node-local heuristics or execution hints.
\end{itemize}
This allows UPC routing to adapt dynamically based on system context.
\paragraph{Programmable Aggregation Logic}
Similarly, how results are aggregated or validated is defined within the Lyquid itself:
\begin{itemize}
    \item Quorum size and validation rules can come from network state (e.g., a global quorum threshold or validator set).
    \item Aggregation reducers (e.g., sum, median, majority) can be programmed explicitly, even using local inputs.
\end{itemize}
\paragraph{Ethereum-Compatible Entrypoints}
Because UPC logic is defined in instance functions, they can be invoked via standard \texttt{eth\_call} interfaces with Ethereum ABI. This means:
\begin{itemize}
    \item Wallets can make the initial call just like ordinary smart contract calls.
    \item Lyquid can internally further extend the execution of this single-point Ethereum RPC into UPC with a complex distributed flow.
    \item No SDK changes needed, just "smarter" contracts.
\end{itemize}
\subsection{Off-Chain Composability}
UPC procedures can call other UPCs, or be composed with other off-chain logic, just like ordinary function calls. This enables powerful coordination patterns such as:
\begin{itemize}
    \item Self-recursive computation across node subsets
    \item Nested off-chain workflows with state-aware control logic
\end{itemize}
In essence, UPC makes your off-chain computation cloud-native, resilient, and composable.


\renewcommand{\refname}{\spacedlowsmallcaps{References}} 

\bibliographystyle{plain}

\bibliography{bibl.bib} 

\appendix
\end{document}